\documentclass[12pt]{article}
\usepackage{amsmath,amssymb,xcolor,pstool,amsthm,mathtools}
\usepackage[unicode,colorlinks = true,
            linkcolor = blue!70!black,
            urlcolor  = red!70!black,
            citecolor = green!55!black,
            anchorcolor = blue!70!black]{hyperref}
\usepackage[top=0.5in, bottom=0.6in, left=0.5in, right=0.5in]{geometry}

\def\di{\mathrm{d}}
\newcommand{\gammae}{\gamma_{\textsf{e}}}
\newcommand{\gammaSP}{\gamma_{\textsf{SP}}}
\newcommand{\tauint}{\tau_{\textsf{int}}}
\newcommand{\tint}{t_{\textsf{int}}}
\newcommand{\tauSP}{\tau_{\textsf{SP}}}
\newcommand{\aRe}{a^{\textsf{Re}}}
\newcommand{\aIm}{a^{\textsf{Im}}}
\newcommand{\phiRe}{\phi^{\textsf{Re}}}
\newcommand{\phiIm}{\phi^{\textsf{Im}}}

\newcommand\blfootnote[1]{
  \begingroup
  \renewcommand\thefootnote{}\footnote{#1}
  \addtocounter{footnote}{-1}
  \endgroup
}

\def\red{\textcolor{red}}
\def\green{\textcolor{black!50!green}}
\def\orange{\textcolor{orange}}

\begin{document}
\begin{titlepage}
\hfill \\
\vspace*{15mm}
\begin{center}
{\LARGE \bf KSW criterion in large field models}

\vspace*{15mm}

{\large Oliver Janssen\blfootnote{e-mail: \href{mailto:oliver.janssen@epfl.ch}{\tt{oliver.janssen@epfl.ch}}}}

\vspace*{8mm}

Laboratory for Theoretical Fundamental Physics, EPFL, 1015 Lausanne, Switzerland

\vspace*{0.7cm}

\end{center}

\begin{abstract}
\noindent \normalsize We extend the analytic description of complex no-boundary solutions in the context of inflation to large field models. We discuss the Kontsevich-Segal-Witten (KSW) criterion and find it is satisfied in small field models, while in large field models it depends on an integral involving $V'(\phi)$ over the range of inflation. It follows that the criterion does not truly constrain inflationary phenomenology since one can complete any inflaton potential beyond observable scales so as to satisfy KSW.
\end{abstract}

\vspace{1cm}
\today

\end{titlepage}

\tableofcontents

\section{Introduction and conclusions}
The no-boundary proposal \cite{PhysRevD.28.2960} is a mathematically compelling candidate for the wave function of a closed universe. It has proven difficult however to reconcile its predictions with observations, as, without further input, it favors histories of the universe with small amounts of inflation with exponential pressure. It appears to predict a value for the spatial curvature parameter which is as large as possible, or, in extremis, an empty de Sitter universe. For reviews about the proposal and possible ways around this substantial unresolved issue see \cite{Maldacena:2024uhs,Lehners:2023yrj,Hartle:2007gi}.

In these notes we assume this hurdle may one day be overcome, and that after the dust has settled the no-boundary geometry is still a potentially relevant contribution to the wave function of the universe in the semiclassical limit. In this context an interesting suggestion was made by Witten \cite{Witten:2021nzp}, building on work in axiomatic quantum field theory by Kontsevich and Segal \cite{Kontsevich:2021dmb}: only complex metrics belonging to the class of ``allowable geometries" can serve as contributing saddles in the semiclassical limit. Colloquially, the allowable geometries are those on which correlation functions of probe matter can be computed by a Euclidean path integral. For diagonal complex metrics $g_{\mu \nu} = \lambda_\mu \delta_{\mu \nu}$ in some basis, the Kontsevich-Segal-Witten (KSW) criterion of allowability can be simply recast into the inequality
\begin{equation} \label{KSWcriterion}
	\sum_{\mu = 0}^{D-1} \left| \arg \lambda_\mu \right| < \pi \,.
\end{equation}
Euclidean metrics lie in the interior of the class of allowable geometries while Lorentzian ones lie on the boundary.

In \cite{Hertog:2023vot}, complex no-boundary solutions that arise in the context of single-field inflation were numerically subjected to the KSW criterion in a collection of models. Intriguingly, the corresponding slow roll histories that passed the criterion were found to have a small tensor-to-scalar ratio $r$ in accordance with observations. This is exciting because $r$ may be our most direct probe of quantum gravity through the CMB, and constraining it on theoretical grounds could explain why we do not observe a large signal. An analytic understanding of this unexpected feature was lacking, however, and in \cite{Maldacena:2024uhs} significant progress was made in this direction. The main point of this paper is to complete the calculation of \cite{Maldacena:2024uhs} to the large field regime (if there is one), where it turns out the criterion \eqref{KSWcriterion} has a chance of failing. To leading order in slow roll we find the KSW criterion translates to
\begin{equation} \label{KSWfinalresult}
	\frac{V'_*}{V_*} \int_{\phi_*}^\chi |\di \phi| \frac{V_*'}{V'(\phi)} < 1 \,,
\end{equation}
where $\phi_*$ is the start of inflation and $\chi$ is the end.


The left-hand side of \eqref{KSWfinalresult} decreases with an increasing amount of efolds if the potential has a hilltop-like shape ($|V'(\phi)|$ decreases with increasing $N_e$). This suggests that \textit{any} potential $V(\phi)$ could be modified beyond the point $\phi_{60}$, corresponding to the regime where observable modes exited the horizon during inflation, so as to satisfy the KSW criterion when the amount of efolds is increased beyond 60. In other words one can engineer the factor $V'_*/V_*$ to decrease dramatically without changing the value of the integral by much. We demonstrate this point in \S\ref{alphasec} and \S\ref{naturalsec} with examples.

Unfortunately this nullifies an ambitious prospect expressed in \cite{Hertog:2023vot} and prohibits KSW from having universal phenomenological implications for inflation (e.g.~small $r$) because we do not probe the inflationary potential beyond $\phi_{60}$, and there is no reason to expect inflation lasted precisely long enough to solve the standard problems in cosmology. One can still contemplate using KSW to rule out any particular slow roll history of the universe, although it is clear that further work on the foundations of the criterion is overdue.\footnote{Examples of complex saddles that do not satisfy KSW but give physically sensible results have been noted in the recent literature, e.g.~\cite{Chen:2023hra}.}

\section{Review of recent literature}
For metric and scalar field
\begin{equation}
	\di s^2 = -\di t^2 + a(t)^2 \di \Omega_3^2 \,, ~~~ \phi(t) \,,
\end{equation}
the EOM are ($1/8 \pi G_N = 1$)
\begin{align}
	\left( \frac{\dot{a}}{a} \right)^2 + \frac{1}{a^2} - \frac{1}{3} \left( \frac{1}{2} \dot{\phi}^2 + V(\phi) \right) &= 0 \,, \label{EOMfull1} \\
	\ddot{\phi} + 3 \frac{\dot{a}}{a}\dot{\phi} + V'(\phi) &= 0 \,. \label{EOMfull2}
\end{align}
In \cite{Maldacena:2024uhs}, an analytic solution to these equations was obtained to first order in slow roll and with no-boundary initial conditions. This reads, in slightly different notation ($\tau = H_* t, H_* = \sqrt{V_*/3}, \varepsilon_* = (V'_*/V_*)^2/2$):

\begin{equation} \label{EOMsol}
\begin{aligned}[c]
\phi(\tau) &= \phi_* + \sqrt{2 \varepsilon_*} \varphi(\tau) \,, \\
\varphi(\tau) &= f_\phi(\tau) + c_\phi \,, \\
f_\phi(\tau) &= \frac{1 + i \sinh \tau}{\cosh^2 \tau} - \log \left( 1 - i \sinh \tau \right) \,,
\end{aligned}
\qquad\qquad
\begin{aligned}[c]
a(\tau) &= \frac{1}{H_*} \cosh(\tau) \left[ 1 + \varepsilon_* \gamma(\tau) \right] \,, \\
\gamma(\tau) &= \tanh \tau \left( f_a(\tau) + c_a \right) \,, \\
f_a(\tau) &= \text{(long explicit expression)} \,.
\end{aligned}
\end{equation}
$c_{\phi,a}$ are integration constants which we have not yet fixed, and $f_a$ depends on $c_\phi$. The center of the ball, or ``south pole" of the geometry, is located near $\tau = i \pi/2$ where
\begin{equation}
	f_\phi(\tau) = \frac{1}{2} - \log 2 + \mathcal{O} \left( \left( \tau - \frac{i \pi}{2} \right)^2 \right) \,, \quad f_a(\tau) = \mathcal{O} \left( \left( \tau - \frac{i \pi}{2} \right)^3 \right) \,.
\end{equation}
For $\tau \gg 1$ on the real line, where the $S^3$ boundary lies,
\begin{align}
	\varphi(\tau) &= - \tau + \log 2 + \frac{i \pi}{2} + c_\phi + 3 e^{-2\tau} - \frac{16i}{3} e^{-3\tau} + \cdots \,, \label{expansionvarphi} \\
	\gamma(\tau) &= - \frac{\tau^2}{2} + \frac{1}{6} \left( 1 + 6 c_\phi + 6 \log 2 + 3 \pi i \right) \tau + \frac{1}{72} \left( 136 - 72 c_\phi - 6 \pi i - 36 \pi i c_\phi - 3\pi^2 - 72 \log 2 - 6 \pi i \log 64 \right) \notag \\
	&+ c_a + (\cdots) e^{-2\tau} + \frac{32i}{9} e^{-3\tau} + \cdots \,. \label{expansiongamma}
\end{align}
To get an approximately real solution on the real line we must set $\text{Im} \, c_\phi = -\pi/2$.\footnote{In any exact no-boundary solution, $a,\phi$ must become exactly real at the same point on the real line. To make this happen, $c_{\phi,a}$ will be corrected by terms exponentially small in $\tau$ and higher order in slow roll.} The real part of $c_\phi$ is redundant with $\phi_*$. As explained in \cite{Maldacena:2024uhs}, setting it to zero causes $\phi_*$ to be identified with the start of inflation in the familiar slow roll solution in flat slicing, viz.~$\phi \sim \phi_* - \sqrt{2 \varepsilon_*} N_e$ at large $N_e \gg 1$ where $N_e = \log a/a_* \sim \tau - \log 2$ is the amount of efolds of inflation and $a_* = 1/H_*$. Likewise, the imaginary part of $c_a$ is a fixed constant dictated by the expansion \eqref{expansiongamma}.\footnote{It depends on $\text{Re} \, c_\phi$. Picking these imaginary parts for $c_{\phi,a}$ causes the leading imaginary terms in \eqref{expansionvarphi}-\eqref{expansiongamma} to be the exponentially small ones $\propto e^{-3\tau}$ that are written.} The real part is redundant with a shift in the time coordinate. Again one could fix it by matching onto the solution in flat slicing. In any case, the south pole is located at
\begin{equation} \label{newSPbis}
	\tauSP = \frac{i \pi}{2} - \varepsilon_* c_a
\end{equation}
including the first slow roll correction. Near this point $a(t) = i \left( t - t_\textsf{SP} \right) + \cdots$ as required by regularity; moving downwards from here, $t = t_\textsf{SP} - i \theta$, $\di s^2 \sim \di \theta^2 + \theta^2 \di \Omega_3^2$. Following the matching condition of \cite{Maldacena:2024uhs}, $c_a$ has both real and imaginary parts. Not following this matching, we could choose $c_a = 0$, causing no shift in the location of the south pole, provided we set $c_\phi = 5/6-\log2 - i \pi/2$. A third possibility is to set $\text{Re } c_a = 0$, as was done in \cite{Janssen:2020pii}. Further, instead of identifying $\phi_*$ with the start of inflation in the flat slicing solution, one could identify $\phi_* = \text{Re } \phi_\textsf{SP}$. We have to first non-trivial order
\begin{equation}
	\phi_\textsf{SP} = \phi_* + \frac{V'_*}{V_*} \left( \frac{1}{2} - \log 2 + c_\phi \right) \,,
\end{equation}
so setting $\text{Re } c_\phi = \log 2 - 1/2$ achieves $\text{Re } \phi_\textsf{SP} = \phi_*$. In turn this sets $\text{Im } c_a = \pi(\log 2 - 2/3)$. In this case the south pole gets shifted by
\begin{equation}
	\tauSP = \frac{i \pi}{2} \left( 1 + \alpha \right)
\end{equation}
where
\begin{equation}
	\alpha = 2 \left( \frac{2}{3} - \log 2 \right) \varepsilon_* = (-0.052961\dots )\varepsilon_* \,.
\end{equation}
This explains analytically what was found approximately in \cite{Janssen:2020pii}, where it was estimated that $\alpha \approx -0.05\varepsilon_*$. The other numerical estimate of \cite{Janssen:2020pii} was $|\text{Im } \phi_\textsf{SP}| \approx 2.21 \sqrt{\varepsilon_*}$. Here we see $2.21$ is really $\pi/\sqrt{2} = (2.2214\dots)$. Finally notice that \eqref{EOMsol} has a singularity ($a,\phi \to \infty)$ at $\tau = i \pi/2 + i(2n+1)\pi$ for integer $n$. In particular there is a singularity at $\tau = -i \pi/2$ at a distance $\pi$ from $\tauSP$. This was anticipated in \cite{Janssen:2020pii} where the first few terms of the series expansion of the solution \eqref{EOMsol} around the south pole were found, and from them it was estimated that the radius of convergence of the series should be about $\pi$.

\section{Regime of validity of \eqref{EOMsol}} \label{validitysec}
The approximation \eqref{EOMsol} is accurate as long as
\begin{equation} \label{smallfieldbound}
	\tau \ll \frac{1}{\sqrt{\varepsilon_*}} \,.
\end{equation}
Beyond this regime, from \eqref{expansiongamma}, \eqref{EOMsol} would predict a negative scale factor. More formally the breakdown can be seen from the equation for $\varphi$,
\begin{equation}
	\ddot{\varphi} + H_* \tanh \tau \, \dot{\varphi} + V'_* = 0 \,,
\end{equation}
where it was assumed that
\begin{equation} \label{adotoveraapprox}
	\frac{\dot{a}}{a} \approx H_* \tanh \tau \,.
\end{equation}
Instead, from \eqref{EOMsol}, keeping only the leading correction term $\propto \varepsilon_* \tau^2$ for $\tau \gg 1$,
\begin{equation}
	\frac{\dot{a}}{a} = H_* \left( \tanh \tau - \frac{2\varepsilon_* \tau}{2-\varepsilon_* \tau^2} \right) \,.
\end{equation}
The approximation \eqref{adotoveraapprox} is not good when $\varepsilon_* \tau^2$ is not small. In other words \eqref{EOMsol} is no longer valid when $\varepsilon_* \gamma = \mathcal{O}(1)$. Another way to put it is that \eqref{EOMsol} is accurate as long as $\Delta \phi \sim \sqrt{\varepsilon_*} \tau \ll 1$. In large field models of inflation this can be violated. Beyond $\tau \gtrsim 1/\sqrt{\varepsilon_*}$ higher order corrections in slow roll $\propto \varepsilon_*^n f(\tau)$ become important.

This last point can be seen explicitly from the usual slow roll solution (in closed or flat slicing), where $a \sim \exp \left( \tau - \varepsilon_* \tau^2/2 \right)$ in the regime $1 \ll \tau \ll 1/\varepsilon_*$. This follows from the relation $H = \dot{a}/a \approx H_*(1-\varepsilon_* \tau)$ in this regime. In large field models such as quadratic inflation $\varepsilon_* \tau^2$ may become large, so in these cases we cannot expand the exponential in the expression for $a$ and keep the first term.

The other approximation that was made to arrive at \eqref{EOMsol} is $V' \approx V'_*$. This is valid as long as
\begin{equation}
	\tau \ll \frac{1}{\eta_*} \,.
\end{equation}

\section{KSW in small field models} \label{smallfieldKSW}
We now turn to the KSW criterion in the small field regime where \eqref{EOMsol} is accurate. This has also been discussed in \cite{Maldacena:2024uhs}. The equation for the extremal curve $t = \gammae(\ell)$, saturating the KSW bound \eqref{KSWcriterion}, reads
\begin{equation}
	\arg \gamma'_\textsf{e} + \arg a(\gammae)^3 = 0 \,.
\end{equation}
This is solved e.g.~by
\begin{equation} \label{tauprimea}
	\gamma'_\textsf{e} = \frac{1}{a^3} \,,
\end{equation}
which in turn is integrated to yield
\begin{equation}
	\int_{\gammaSP}^{\gammae} \di \gamma \, a(\gamma)^3 = \ell \,,
\end{equation}
or, taking the imaginary part\footnote{A parametrization of $\gammae$ has implicitly been chosen by the solution \eqref{tauprimea}, but this choice has been removed by taking the imaginary part in \eqref{KSWcontourEq}.},
\begin{equation} \label{KSWcontourEq}
	\text{Im} \left( \int_{\gammaSP}^{\gammae} \di \gamma \, a(\gamma)^3 \right) = 0 \,.
\end{equation}
We can compute the integral by splitting the contour into 3 pieces: calling $I = I_{\gammaSP \to \gammae} \equiv \int_{\gammaSP}^{\gammae} \di \gamma \, a(\gamma)^3$ we write
\begin{equation}
	I = I_{\gammaSP \to 0} + I_{0 \to u} + I_{u \to u + i \sigma} \,,
\end{equation}
where $u + i \sigma = \gammae$. The dominant contribution to the imaginary part of the first piece comes from the non-$\varepsilon_*$-suppressed part of $a$, that is, the de Sitter solution $a(\tau) = (1/H_*) \cosh \tau$. We have
\begin{equation}
	\text{Im} \left( I_{\gammaSP \to 0} \right) = -\frac{2}{3H_*^4} \left[ 1 + \mathcal{O}\left( \varepsilon_* \right) \right] \,.
\end{equation}
Then,
\begin{equation}
	I_{0 \to u} = \text{(real)} + \frac{3 \varepsilon_*}{H_*^3} \int_0^u \di t \cosh^3(H_* t) \gamma(t) + \cdots \,.
\end{equation}
We have $\text{Im }\gamma(t) \approx (32/9) e^{-3H_*t}$ after a time $t \gtrsim 1/H_*$ when slow roll kicks in. Therefore we can split the integral into a piece $0 \to \mathcal{O}(1/H_*)$ giving a small contribution $(\varepsilon_*/H_*^4) \times \mathcal{O}(1)$ plus the contribution
\begin{equation}
	\text{Im}(I_{0 \to u}) = \mathcal{O}\left( \frac{\varepsilon_*}{H_*^4} \right) + \frac{3 \varepsilon_*}{H_*^3} \int_{\mathcal{O}(1)}^u \di t \, \frac{1}{8} e^{3H_* t} \frac{32}{9} e^{-3H_* t} = \frac{4\varepsilon_*}{3H_*^4} \times  \left[ H_* u + \mathcal{O}(1) \right]
\end{equation}
for $H_* u \gg 1$. For the last piece
\begin{equation}
	\text{Im}\left( I_{u \to u + i \sigma} \right) \approx \frac{e^{3 H_* u}}{8 H_*^4} H_* \sigma \,.
\end{equation}
Putting it all together in \eqref{KSWcontourEq}, solving for $\sigma$ and calling $H_* u = \tau$,
\begin{equation} \label{Maldabound}
	H_* \sigma = \frac{16}{3} e^{-3\tau} \left[ 1 + \mathcal{O}(\varepsilon_*) - 2 \varepsilon_* \tau \right] \,,
\end{equation}
which was obtained in \cite{Maldacena:2024uhs}.

We argued in \S\ref{validitysec} that the approximation \eqref{EOMsol}, which was used in this calculation, is valid as long as $\sqrt{\varepsilon_*} \tau \ll 1$. So surely in this case $\varepsilon_* \tau \ll 1$ and the extremal curve remains above the real axis ($\sigma > 0$ in \eqref{Maldabound}), where the boundary lies. This means we can construct a KSW-allowable curve that reaches the boundary. We conclude that small field models satisfy the KSW criterion.\footnote{Perhaps this is not quite rigorous. We have argued that if $\sqrt{\varepsilon_*} \tau \ll 1$, then there is a small (sub-Planckian) field variation and KSW is satisfied. For the converse, standard manipulations yield $\Delta \phi = \int \di t \, H\sqrt{2 \varepsilon_H} \gtrsim \sqrt{2 \varepsilon_*} \int \di t \, H \geq \sqrt{2 \varepsilon_*} \tau (H_\textsf{end}/H_*)$, where we have used $\varepsilon_H = \varepsilon_V$ to first order in slow roll (for a mini review of slow roll see \S\ref{largefieldsec}). So if $\Delta \phi \ll H_\textsf{end}/H_*$, which is perhaps slightly stronger than small field ($\ll 1$), KSW would be satisfied. Our final result is really Eq. \eqref{KSWboundNEW} below which makes no independent reference to the distance traversed by $\phi$.}

Formula \eqref{Maldabound} suggests that KSW could be violated when $\varepsilon_* \tau$ becomes of order one, as it does in some large field models of inflation e.g.~monomial inflation where $V = \phi^p$ and $\varepsilon_* \tau \sim p/4$ for large $\phi_* \gg \chi$, where $\chi$ is the end of inflation. This is not sharp, however, because the approximation used to arrive at \eqref{Maldabound} is not valid in this regime. In the next section we extend the approximate solution \eqref{EOMsol} beyond $\sqrt{\varepsilon_*} \tau \ll 1$ to the whole slow roll patch and then discuss the KSW criterion in the extended regime.

\section{Approximate solution in large field models} \label{largefieldsec}
To extend the solution \eqref{EOMsol} beyond $\tau \ll 1/\sqrt{\varepsilon_*}$, we reconsider the EOM \eqref{EOMfull1}-\eqref{EOMfull2} with the scale factor and scalar field written as a dominant real part plus a small imaginary part:
\begin{equation}
	a = \aRe + i \aIm \,, \quad \phi = \phiRe + i \phiIm \,.
\end{equation}
Linearizing in the imaginary parts, the zeroth order equations are simply \eqref{EOMfull1}-\eqref{EOMfull2} with $(a,\phi) \to (\aRe,\phiRe)$. For $\tau \gg 1$ these become the usual slow roll equations \cite{Kolb:1994ur,Liddle:1994dx}, about which we recall some useful formulas. Neglecting the curvature term altogether in what follows\footnote{This is reasonable because this term is volume-suppressed compared to others, which may only be slow roll-suppressed.}, the EOM for the real parts can be put in the otherwise exact form
\begin{align}
	3 H^2 = \frac{(\dot{\phi}^\textsf{Re})^2}{2} + V(\phiRe) \,, \quad (3-\eta_H) H \dot{\phi}^\textsf{Re} = -V'(\phiRe) \,, \label{EOMH1}
\end{align}
where
\begin{equation} \label{SRdefs}
	H \equiv \frac{\dot{a}^\textsf{Re}}{\aRe} \,, \quad \eta_H \equiv - \frac{\ddot{\phi}^\textsf{Re}}{H \dot{\phi}^\textsf{Re}} \,.
\end{equation}
Is it also useful to define
\begin{equation}
	\varepsilon_H \equiv - \frac{\dot{H}}{H^2} \,, \quad \zeta_H^2 \equiv \frac{1}{H^2} \partial_t (\ddot{\phi}/\dot{\phi}) \,.
\end{equation}
One can derive the (exact) relations
\begin{align}
	V &= (3-\varepsilon_H)H^2 \,,
& V' &= \sqrt{2\varepsilon_H}(3-\eta_H)H^2 \,,
& V'' &= (3\varepsilon_H + 3\eta_H - \eta_H^2 - \zeta_H^2)H^2 \,, \\
\varepsilon_H &= \frac{(\dot{\phi}^\textsf{Re})^2}{2 H^2} \,,
& \ddot{a}^\textsf{Re} &= \aRe (1-\varepsilon_H)H^2 \,. \label{EOMH2}
\end{align}
Although we will not use them, for completeness we recall relations between the Hubble slow roll parameters $\varepsilon_H, \eta_H, \zeta_H$ and the potential slow roll parameters $\varepsilon_V, \eta_V,\zeta_V$, defined as
\begin{equation}
	\varepsilon_V = \frac{1}{2} \left( \frac{V'}{V} \right)^2 \,, \quad  \eta_V = \frac{V''}{V} \,, \quad \zeta_V^2 = \frac{V' V'''}{V^2} \,,
\end{equation}
To second order \cite{Kolb:1994ur},
\begin{equation}
	\varepsilon_H = \varepsilon_V - \frac{4}{3} \varepsilon_V^2 + \frac{2}{3} \varepsilon_V \eta_V + \cdots \,, \quad \eta_H = \eta_V - \varepsilon_V + \frac{8}{3} \varepsilon_V^2 - \frac{8}{3} \varepsilon_V \eta_V + \frac{1}{3} \eta_V^2 + \frac{1}{3} \zeta_V^2 + \cdots \,.
\end{equation}

Returning to the EOM, the first order equations (or, leading imaginary parts) are \cite{Janssen:2020pii}
\begin{align}
	H \dot{a}^\textsf{Im} - \frac{\aRe}{6} \left( V'(\phiRe) \phiIm + \dot{\phi}^\textsf{Re} \dot{\phi}^\textsf{Im} \right) - H^2 \aIm = \frac{\aIm}{(\aRe)^2} &\approx 0 \,, \label{EOMapproxIm1T} \\
	\ddot{\phi}^\textsf{Im} + 3H \dot{\phi}^\textsf{Im} + 3 \frac{\dot{\phi}^\textsf{Re}}{\aRe} \left( \dot{a}^\textsf{Im} - H \aIm \right) + V''(\phiRe) \phi^\textsf{Im} &= 0 \,. \label{EOMapproxIm2T}
\end{align}
In \eqref{EOMapproxIm1T} we have again neglected a term which is volume-suppressed compared to the others. Inspired by \eqref{EOMsol}, which we must reproduce when $1 \ll \tau \ll 1/\sqrt{\varepsilon_*}$, we try the Ansatz
\begin{equation} \label{Ansatzsubleading}
	\phiIm = \frac{g_\phi}{(\aRe)^3} \,, \quad \aIm = \frac{g_a}{(\aRe)^2} \,,
\end{equation}
where $g_{\phi,a}$ are slowly-varying functions, i.e.~the dimensionless $\dot{g}/(Hg)$ are slow roll-suppressed. In \eqref{EOMsol} these functions are just constants. In this way
\begin{equation}
	\dot{\phi}^\textsf{Im} = -3H \phiIm \left( 1 - \frac{1}{3} \frac{\dot{g_\phi}}{H g_\phi} \right) \,, \quad \dot{a}^\textsf{Im} = -2H \aIm \left( 1 - \frac{1}{2} \frac{\dot{g_a}}{H g_a} \right) \,,
\end{equation}
where the correction terms are small. More precisely we will set up a perturbative expansion
\begin{align}
	g_\phi = g_\phi^{(0)} + g_\phi^{(1)} + \cdots \,, \quad g_a = g_a^{(0)} + g_a^{(1)} + \cdots \,,
\end{align}
treating $g^{(i+1)} \ll g^{(i)}$ but $g^{(i+1)} = \mathcal{O} \left( \dot{g}^{(i)}/H g^{(i)} \right)$.

 Plugging \eqref{Ansatzsubleading} into \eqref{EOMapproxIm1T}, using \eqref{EOMH1}-\eqref{EOMH2}, gives
\begin{align} \label{146}
	\green{g_a^{(0)} + \frac{\sqrt{2 \varepsilon_H}}{3} g_\phi^{(0)}} \orange{ + g_a^{(1)} + \frac{\sqrt{2 \varepsilon_H}}{3} g_\phi^{(1)} - \frac{\sqrt{\varepsilon_H}}{9 \sqrt{2}} \left( \eta_H + \frac{\dot{g}_\phi^{(0)}}{H g_\phi^{(0)}} \right)g_\phi^{(0)} - \frac{\dot{g}_a^{(0)}}{3H}} \red{+ \cdots} = 0 \,,
\end{align}
where leading terms are indicated in green, subleading ones in orange, and we have neglected even further subleading terms in red. Solving the leading piece gives
\begin{equation} \label{29leading}
	\green{g_a^{(0)} = - \frac{\sqrt{2 \varepsilon_H}}{3} g_\phi^{(0)}} \,.
\end{equation}

Plugging \eqref{Ansatzsubleading} into \eqref{EOMapproxIm2T} yields instead
\begin{align}
	&\green{3 \sqrt{2 \varepsilon_H} \, g_a^{(0)} + (2\varepsilon_H + \eta_H) g_\phi^{(0)} - \frac{\dot{g}_\phi^{(0)}}{H}}   \\
	&\orange{+3 \sqrt{2 \varepsilon_H} g_a^{(1)} - \frac{1}{3} \left( \eta_H^2 + \zeta_H^2 \right) g_\phi^{(0)} + (2\varepsilon_H + \eta_H) g_\phi^{(1)} - \sqrt{2 \varepsilon_H} \frac{\dot{g}_a^{(0)}}{H} - \frac{\dot{g}_\phi^{(1)}}{H} + \frac{\ddot{g}_\phi^{(0)}}{3H^2}} \red{+ \cdots} = 0 \,.\label{subL43}
\end{align}
Considering the leading equation, using \eqref{29leading}, we get
\begin{equation}
	\green{\frac{\dot{g}_\phi^{(0)}}{H g_\phi^{(0)}} = \eta_H} \,,
\end{equation}
which, as promised, is slow roll-suppressed. From the definition of $\eta_H$ in \eqref{SRdefs} we see the solution is
\begin{equation}
	g_\phi^{(0)} = \frac{C}{\dot{\phi}^\textsf{Re}} = - \frac{C}{\sqrt{2 \varepsilon_H} H} \,,
\end{equation}
for a constant $C$ (we assume $V'(\phiRe) > 0$.)
It follows that
\begin{equation}
	g_a^{(0)} = \frac{C}{3H} \,.
\end{equation}
$C$ is determined by matching onto the intermediate $\tau$ solution of \cite{Maldacena:2024uhs} in 
\eqref{EOMsol} where $H, \varepsilon_H \approx H_*, \varepsilon_*$:
\begin{equation} \label{Ceq}
	C \approx \frac{4\varepsilon_*}{3H_*^2} \,.
\end{equation}
We summarize the leading behavior:
\begin{equation} \label{solutionX}
	\phiIm = \frac{g_\phi^{(0)} + \cdots}{(\aRe)^3} = -\frac{C}{\sqrt{2 \varepsilon_H} H (\aRe)^3} \left( 1 + \cdots \right) \,, \quad \aIm = \frac{g_a^{(0)} + \cdots}{(\aRe)^2} = \frac{C}{3 H (\aRe)^2} \left( 1 + \cdots \right) \,.
\end{equation}
We stress that $H$ and $\varepsilon_H$ are slowly-varying functions (their relative change per efold is small), but varying nonetheless. Their variation over the entirety of inflation may be large.

\eqref{solutionX} should be a good approximation as long as slow roll persists. So if slow roll lasts longer than $\tau \sim 1/\sqrt{\varepsilon_*}$, it extends the approximation \eqref{EOMsol} which was valid only for $\tau \ll 1/\sqrt{\varepsilon_*}$. In terms of $\varepsilon_*$, in a sense \eqref{solutionX} involves all powers $\varepsilon_*^n$, as it depends on the functions $H, \varepsilon_H$ which depend on the entire function $V(\phiRe)$. We are familiar with this feature in slow roll, e.g.~the amount of efolds $N_e$ depends on an integral of the function $1/\sqrt{2\varepsilon}$, which is slowly-varying, but its total variation may be large.

Finally we solve the subleading orange equations, which are unimportant for the main point of this paper (the KSW criterion in \S\ref{KSWextendedsec}), but important to analyze potential subleading corrections to the on-shell action coming from the large field regime. These are discussed in Appendix \ref{OSAappendix}.

The subleading equation of \eqref{146} sets
\begin{equation}
	\orange{g_a^{(1)} = \frac{C(\varepsilon_H - \eta_H)}{9H} - \frac{\sqrt{2 \varepsilon_H}}{3} g_\phi^{(1)}} \,.
\end{equation}
We use this in the subleading equation \eqref{subL43}. A useful relation is
\begin{equation}
	\dot{\eta}_H = H(\varepsilon_H \eta_H - \zeta_H^2) \,, \quad \text{so } \ddot{g}_\phi^{(0)} = \left( \zeta_H^2 - \eta_H^2 \right) \frac{C H}{\sqrt{2 \varepsilon_H}} \,.
\end{equation}
We obtain
\begin{equation}
	\orange{\frac{\dot{g}_\phi^{(1)}}{H} = \eta_H g_\phi^{(1)} + \frac{2C}{3 \sqrt{2 \varepsilon_H} H} \left( \zeta_H^2 - \varepsilon_H \eta_H \right)} \,.
\end{equation}
The solution is
\begin{equation}
	g_\phi^{(1)} = \frac{2 \eta_H}{3} g_\phi^{(0)} \,,
\end{equation}
and so
\begin{equation}
	g_a^{(1)} = \frac{1}{3} \left( \varepsilon_H + \eta_H \right) g_a^{(0)} \,.
\end{equation}
Summarizing the subleading behavior:
\begin{equation} \label{solutionX2}
	\phiIm = -\frac{C}{\sqrt{2 \varepsilon_H} H (\aRe)^3} \left( 1 + \frac{2}{3} \eta_H + \cdots \right) \,, \quad \aIm = \frac{C}{3 H (\aRe)^2} \left( 1 + \frac{1}{3} \left( \varepsilon_H + \eta_H \right) + \cdots \right) \,,
\end{equation}
with $C$ given in \eqref{Ceq}.

Before turning to KSW, we note that \eqref{solutionX2} corrects two formulas that appeared in \cite{Janssen:2020pii}, where it was claimed that
\begin{equation} \label{incorrectJanssen}
	\phiIm \approx -\frac{\tilde{C}}{\sqrt{2} H (\aRe)^3} \,, \quad \aIm \approx \frac{\tilde{C} \sqrt{\varepsilon}}{3 H (\aRe)^2} \quad \quad \quad \text{[incorrect for $\tau \gtrsim 1/\sqrt{\varepsilon_*}$]}
\end{equation}
at large $\tau \gg 1$, for a constant $\tilde{C}$. From \eqref{solutionX2} one sees that the leading slowly-varying factors are inaccurate. As long as $\tau \ll 1/\sqrt{\varepsilon_*}$, we have $H, \varepsilon_H \approx H_*, \varepsilon_*$ and \eqref{incorrectJanssen} agrees with the solution \eqref{EOMsol} of \cite{Maldacena:2024uhs} which is valid in this regime. Outside of this regime \eqref{incorrectJanssen} is inaccurate.

\section{KSW in the extended region} \label{KSWextendedsec}
Following the steps of \S\ref{smallfieldKSW} we now use \eqref{solutionX} to calculate the extremal curve. Instead of \eqref{Maldabound}, after a straightforward calculation involving the standard change of variables $t \to \phi$ in slow roll, we obtain
\begin{equation} \label{newKSW2}
	H_* \sigma = \frac{2}{3 (b H_*)^3} \left( 1 + \mathcal{O}(\varepsilon_*) - \sqrt{2 \varepsilon_*} \int_{\phi_*}^\chi |\di \phi| \frac{V_*'}{V'(\phi)} \right) \,,
\end{equation}
where we have indicated the final (real) values of the scale factor and scalar field by $b$ and $\chi$ respectively, and we have also used that $\dot{\phi} = -V'/3H$ to leading order in slow roll. This reduces to \eqref{Maldabound} for $\tau \ll 1/\sqrt{\varepsilon_*}, 1/\eta_*$ since there $V'(\phi) \approx V'_*$ and $\Delta \phi \approx \sqrt{2 \varepsilon_*} \tau$. It is also valid beyond this regime, however, as long as we are in slow roll ($\varepsilon$ and $\eta$ are small in the region between $\phi_*$ and $\chi$). We conclude KSW will be satisfied as long as
\begin{equation} \label{KSWboundNEW}
	\mathcal{A}(\phi_*,\chi) \equiv \frac{V'_*}{V_*} \int_{\phi_*}^\chi |\di \phi| \frac{V_*'}{V'(\phi)} < 1 + \mathcal{O}(\varepsilon_*) \,.
\end{equation}
By $|\di \phi|$ we mean the overall sign should be chosen so that $\mathcal{A}$ is positive. We have neglected an $\mathcal{O(\varepsilon_*})$ correction to the right-hand side of the bound here. This corresponds to a fuzziness of order $\sqrt{\varepsilon_*}$ in what we mean by $\phi_*$, or equivalently a change by an $\mathcal{O}(1)$ amount of efolds of inflation.

For large field models in which $|V'_*|$ is decreasing with increasing $|\phi_*-\chi|$ (or, increasing amount of efolds) at fixed endpoint $\chi$, i.e.~plateaux or hilltop potentials, this bound is qualitatively different from the naive extension to $\tau = \mathcal{O}(1/\varepsilon_*)$ of the small field bound $2 \varepsilon_* \tau < 1$ extrapolated from \eqref{Maldabound}. This is because $\mathcal{A}$ could \textit{decrease} with an increasing amount of efolds, whereas $2 \varepsilon_* \tau \sim 2 \varepsilon_* N_e$ always increases. We will see explicit instances of this in \S\ref{alphasec} and \S\ref{naturalsec} below.

\section{Numerical checks} \label{numsec}

\subsection{Monomial inflation} \label{monomialsec}
For monomial inflation,
\begin{equation}
	V = \phi^p
\end{equation}
with $\phi,p > 0$, we find
\begin{align}
	\frac{V'_*}{V_*} \int_{\phi_*}^\chi |\di \phi| \frac{V_*'}{V'(\phi)} =
	 \begin{dcases}
      2 \, \log \left( \frac{\phi_*}{\chi} \right) & \text{if $p = 2$,} \\
      \frac{p}{2-p} \left[ 1 - \left(\frac{\chi}{\phi_*}\right)^{2-p} \right] & \text{otherwise.}
    \end{dcases}
\end{align}
Setting this to one and solving for $\phi_*$ gives us the value $\phi_{*,\textsf{KSW}}(\chi)$ beyond which KSW will be violated:
\begin{equation} \label{KSWmonomial}
	\phi_{*,\textsf{KSW}}(\chi) = \begin{dcases}
      \sqrt{e} \, \chi & \text{if $p = 2$,} \\
      \frac{\chi}{\left( 2 - \frac{2}{p} \right)^{\frac{1}{2-p}}} & \text{if $p > 1$,} \\
      \infty & \text{if $p \leq 1$.}
    \end{dcases}
\end{equation}
That is, if we start rolling from $\phi_* > \phi_{*,\textsf{KSW}}(\chi)$ and end at $\chi$, the trajectory will not be allowable, while rolling from $\phi_* < \phi_{*,\textsf{KSW}}(\chi)$ and ending at $\chi$ is allowable. The structure is this way -- from non-allowable at small $\phi_* \approx \chi$ to allowable at large $\phi_* \gg \chi$ -- because $\mathcal{A}$ is monotonically increasing with $\phi_*$ at fixed $\chi$.

\begin{figure}[h!]
\centering
\includegraphics[width=340pt]{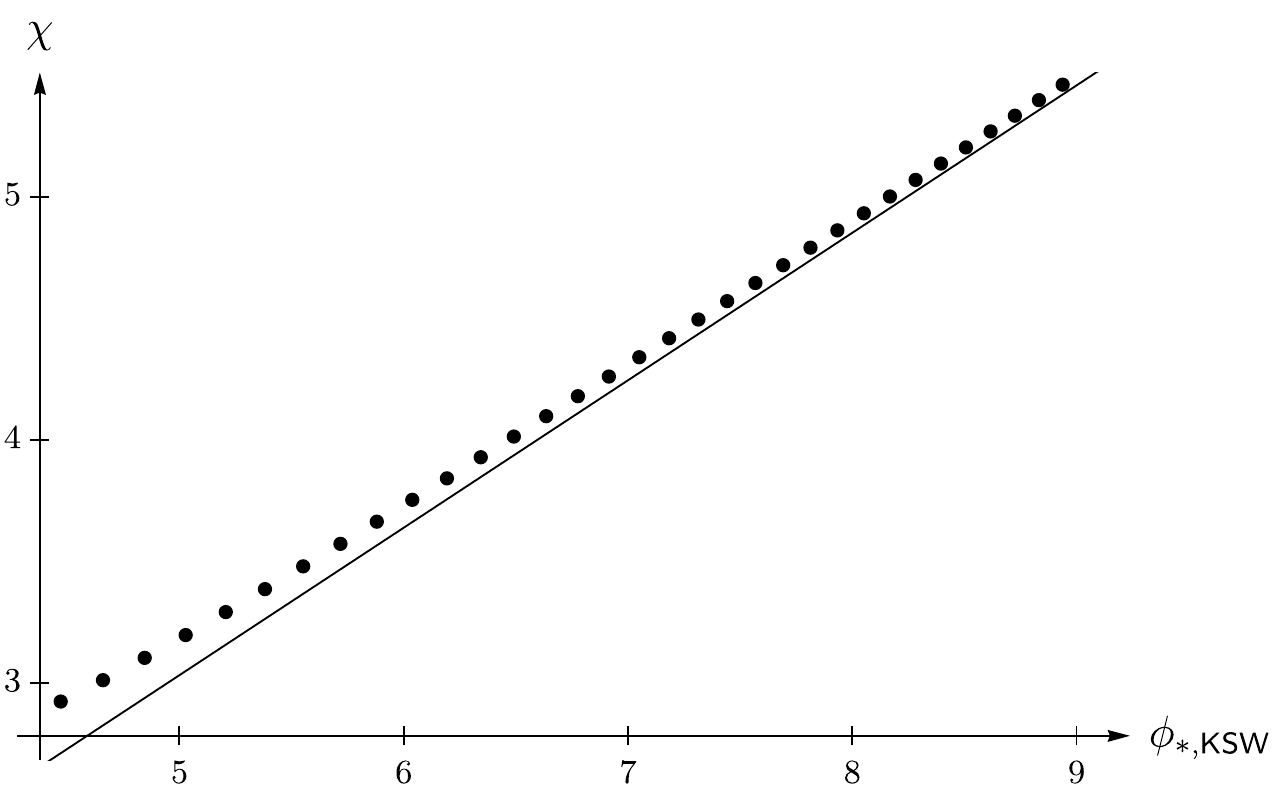}
\caption{Results in $m^2 \phi^2$ inflation ($m$ drops out in slow roll): for a given endpoint of inflation $\chi$, $\phi_{*,\textsf{KSW}}$ is the maximal starting point of inflation so that the slow roll history from $\phi_{*,\textsf{KSW}}$ to $\chi$ is KSW-allowable. Dots are numerical data, the line is the analytic prediction $\phi_{*,\textsf{KSW}} = \sqrt{e} \chi$. Figure courtesy of J. Karlsson.}
\label{FIGm2phi2results}
\end{figure}

For $p = 2$ we have $\chi = e^{-1/2} \phi_{*,\textsf{KSW}}(\chi) \approx (0.606 \dots) \phi_{*,\textsf{KSW}}(\chi)$. This value is consistent with numerics, see Figure \ref{FIGm2phi2results}. Notice that our approximations should be trustworthy in this case since both $\chi$ and $\phi_*$ lie parametrically far in the slow roll region when $\phi_* \gg 1$.
This result can be contrasted with a naive extension of the small field KSW bound $2\varepsilon_* \tau < 1$ inferred from \eqref{Maldabound}, which would not rule out any trajectory since
\begin{equation}
	2\varepsilon_* \tau \sim 2 \frac{p^2}{2\phi_*^2} \times \frac{1}{2p} \left( \phi_*^2 - \chi^2 \right) = \frac{p}{2} \left[ 1 - \left( \frac{\chi}{\phi_*} \right)^2 \right] \,,
\end{equation}
which is smaller than $1$ (and so the trajectory would be allowable) for any $p \leq 2$ and $\chi < \phi_*$.

For $p \leq 1$, $\phi_{*,\textsf{KSW}}(\chi) = \infty$ for all $\chi$ and so all trajectories are allowable according to \eqref{KSWboundNEW}. For $p > 1$ there is an upper bound on the amount of efolds of inflation that can reach a given endpoint $\chi$. This is
\begin{equation} \label{NemaxKSW}
	N_{e,\textsf{max}}(\chi) = \frac{1}{2p} \left[ \frac{1}{\left( 2 - \frac{2}{p} \right)^{\frac{2}{2-p}}} - 1 \right] \chi^2 \,.
\end{equation}
As $p \to 1^+$ this diverges.

Strictly speaking, as we have stressed before, these formulas are valid only in slow roll. In particular $\chi$ must lie in the slow roll regime. In \cite{Hertog:2023vot}, trajectories were examined which end at the exit of slow roll where the slow roll parameters become order one. Therefore we should not expect the above analysis to be exactly correct in that case, but we can proceed with the idea that most of the slow roll history should be described as above. In particular, it was found numerically in \cite{Hertog:2023vot} that $N_{e,\textsf{max}}(\chi = \phi_{\textsf{end}}) = 60$ for $p \approx 1.05.$ According to the analytic estimate \eqref{NemaxKSW}, setting $ \chi = \phi_\textsf{end} = p/\sqrt{2}$, $N_{e,\textsf{max}} = 60$ and solving for $p$ yields
\begin{equation}
	p_{\textsf{KSW},60} = (1.03773\dots) \,,
\end{equation}
which is close.

Finally we have gathered one more isolated -- rather arbitrary -- data point for $p = 3/2$. In this case we set $\phi_* = 8$, for which \eqref{KSWmonomial} predicts inflation should be allowable until $\chi = (3.55\dots)$, or for $N_e \approx 17.12$ efolds. We have $\varepsilon_\chi \approx 0.09$ so we are approximately in slow roll throughout the trajectory from $\phi_*$ to $\chi$. Numerically we found $3.530 < \chi_\textsf{KSW} < 3.534$. Some discrepancy is to be expected here since in \eqref{KSWboundNEW} we have neglected an $\mathcal{O}(\varepsilon_*)$ correction to the right-hand side, corresponding to a change of $\phi_*$ by an amount $\mathcal{O}(\sqrt{\varepsilon_*} \sim 0.13)$.

\subsection{$\alpha$-attractors} \label{alphasec}
\begin{figure}[h!]
\centering
\includegraphics[width=12cm]{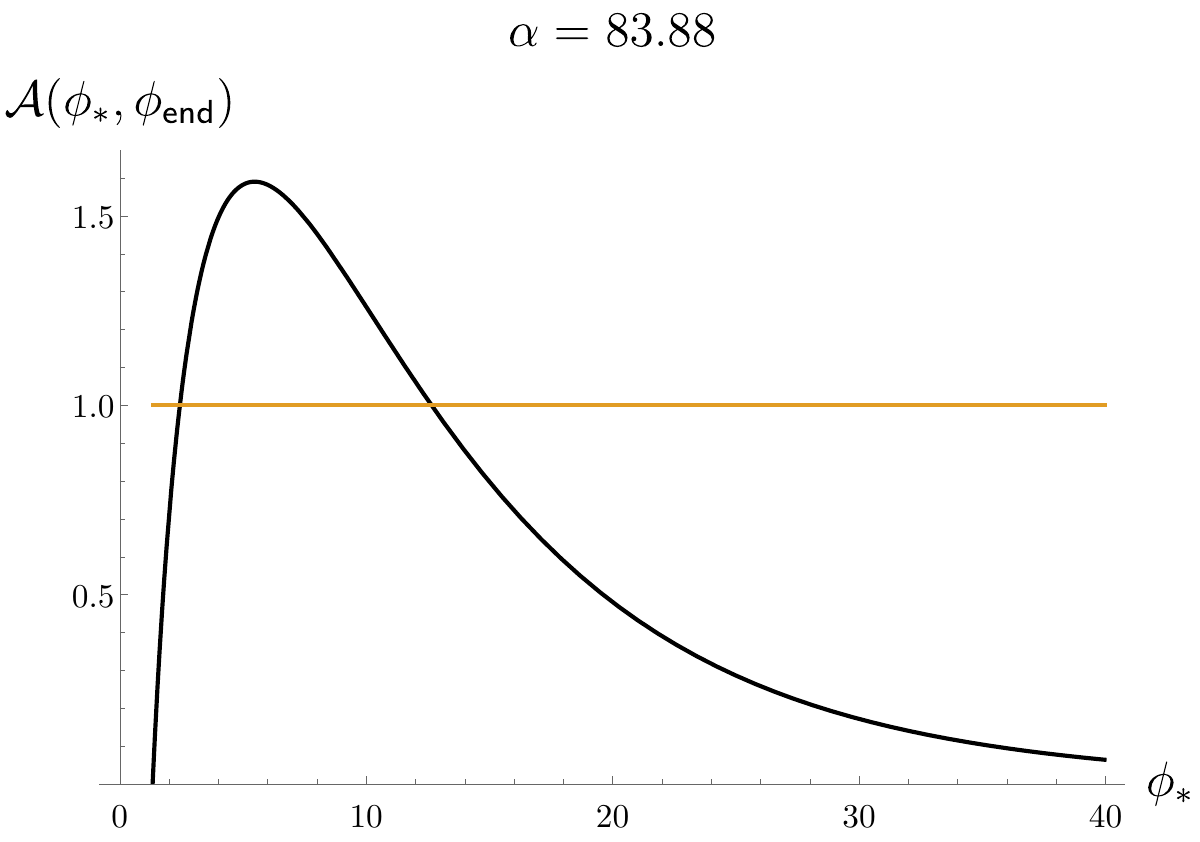}
\caption{The integral $\mathcal{A}$ of Eq. \eqref{KSWboundNEW} for $\alpha$-attractor inflation with $\alpha = 83.88$, which must be smaller than one for KSW to be satisfied. Here we have chosen $\chi = \phi_\textsf{end} = \sqrt{3\alpha/2} \, \log \left( 1 + 2/\sqrt{3\alpha} \right)$ at the end of inflation where $\varepsilon_V = 1$. We see that starting inflation beyond $\phi_* \approx 12.65$, which produces 60 efolds of inflation, is predicted to be allowable, while starting at smaller values is not allowable, unless the amount of efolds is very small (in this case $\approx$ 1.14 efolds; intuitively this regime is KSW-allowable because the geometry is approximately Euclidean). This analytic result is not in poor agreement with numerics (which tell us $\alpha_{\textsf{KSW},60} \approx 93.9$ instead of 83.88), and as we have stressed some disagreement is to be expected since our analytic approximation is not valid towards the end of inflation.}
\label{FIGalpharesults}
\end{figure}

Here
\begin{equation}
	V(\phi) = \left[ 1 - \exp \left( -\sqrt{2}\phi/\sqrt{3 \alpha} \right) \right]^2 \,.
\end{equation}
As for the monomials we can estimate analytically what the value $\alpha_{\textsf{KSW},60}$ should be, which in \cite{Hertog:2023vot} was found to be $\approx 93.9$. In this case the structure is opposite to that of monomial inflation: at fixed $\chi$, $\mathcal{A}$ initially increases with $\phi_*$ for $\phi_* \gtrsim \chi$, but then when $\phi_* \gg \chi$ it decreases again, tending to zero as $\phi_* \to \infty$. This is because here $V'_*$ decreases with increasing $\phi_*$, in contrast to monomial inflation where $V'_*$ grows with $\phi_*$. So here if $\phi_* > \phi_{*,\textsf{KSW}}(\chi)$ (more inflation) the trajectory is allowable while if $\phi_* < \phi_{*,\textsf{KSW}}(\chi)$ (less inflation) it is not allowable, unless $\phi_* \approx \chi$ in which case it becomes allowable again. We leave out the explicit expressions and just state the result $\alpha_{\textsf{KSW},60} = (83.89\dots)$, which is not too far off. We summarize these findings in Figure \ref{FIGalpharesults}.

\begin{figure}[h!]
\centering
\includegraphics[width=12cm]{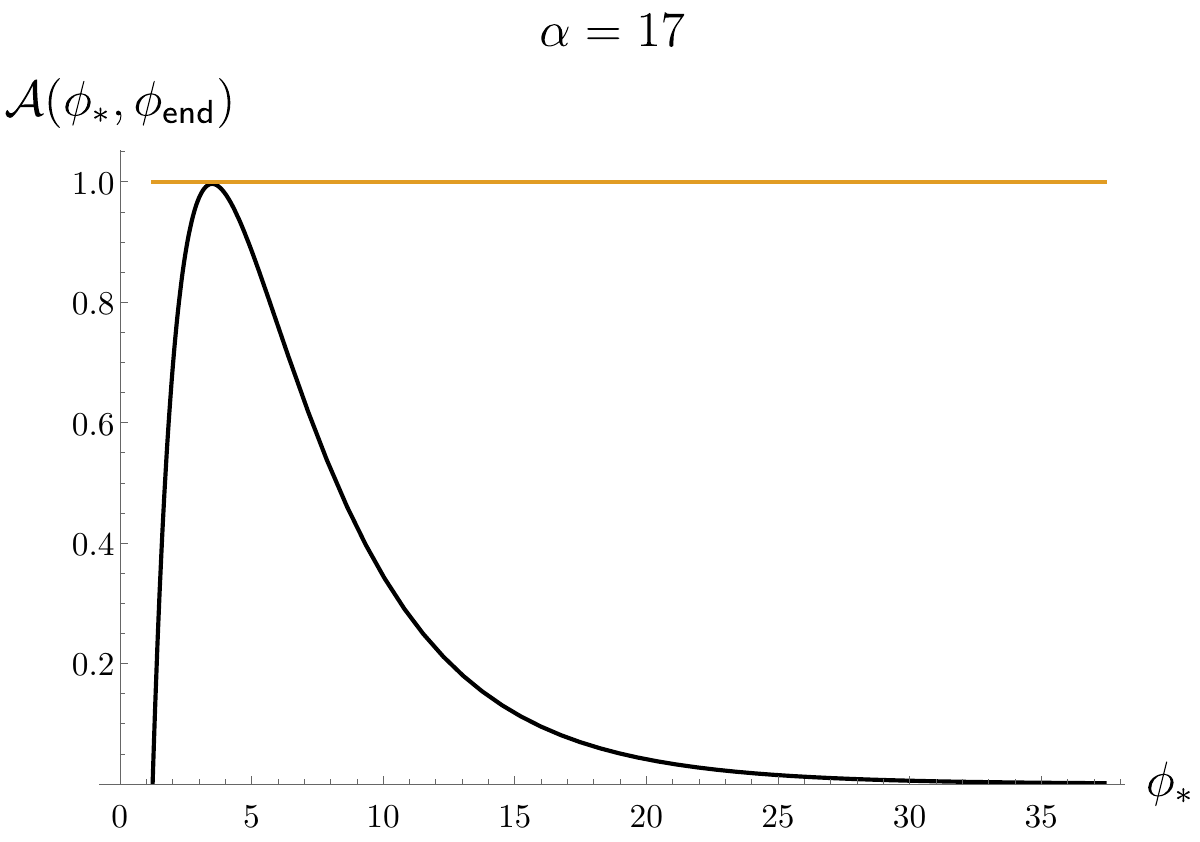}
\caption{$\alpha \approx 17$ is the critical value beyond which KSW begins to rule out some inflationary histories. For smaller $\alpha$ all trajectories (with any amount of inflation) are allowable.}
\label{alphacritFIG}
\end{figure}

It is also curious to note that $\alpha \approx 17$ is the critical value below which all slow roll trajectories are allowable. For larger $\alpha$, like in Figure \ref{FIGalpharesults}, there is some exclusion. We show this in Figure \ref{alphacritFIG}.

\subsection{Natural inflation} \label{naturalsec}
\begin{figure}[h!]
\centering
\includegraphics[width=12cm]{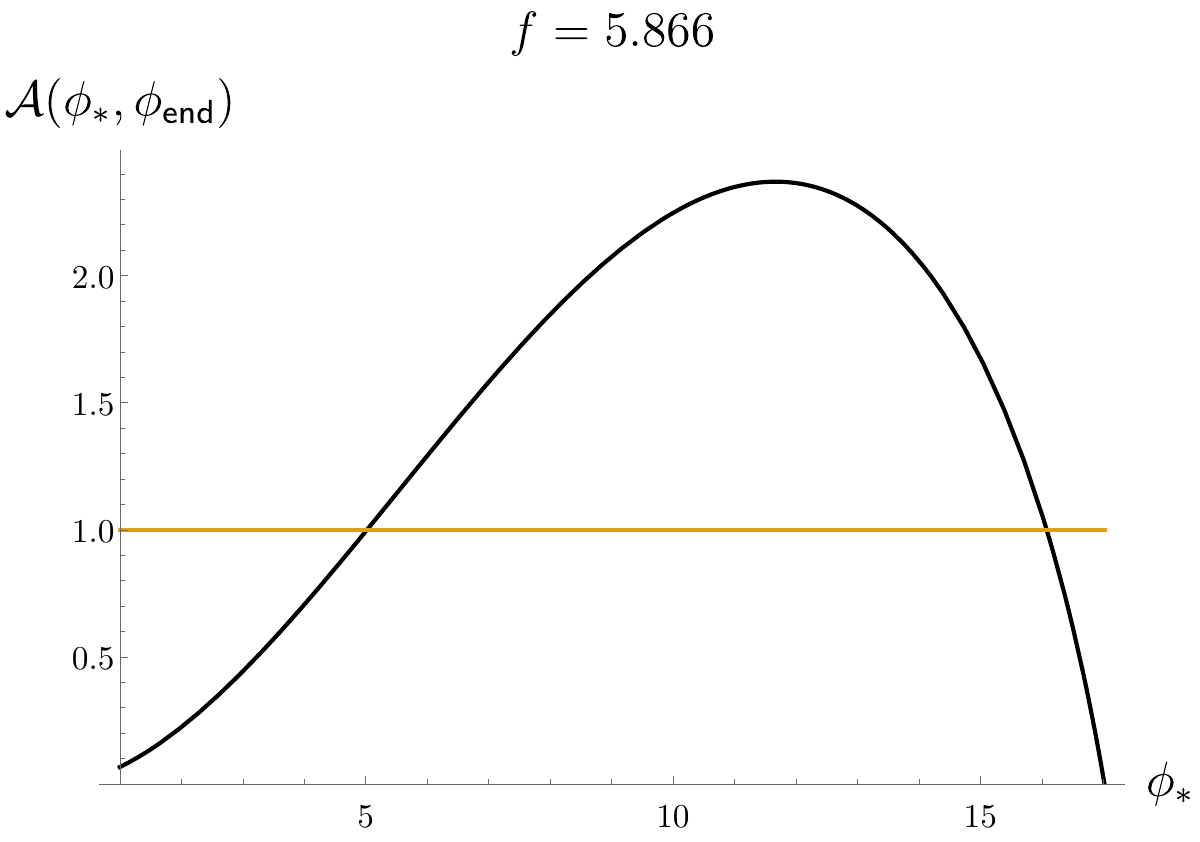}
\caption{Same setup as in Figure \ref{FIGalpharesults}, but now for natural inflation where $V = 1 + \cos(\phi/f)$. Inflation starts near the hilltop at $\phi \approx 0$ and ends near the minimum at $\phi \approx \pi f$. More precisely $\phi_\textsf{end} = 2f \tan^{-1}(\sqrt{2}f)$ is where $\varepsilon_V = 1$. Shown here again is the KSW integral \eqref{KSWboundNEW} which must be smaller than one for an allowable history. We see in this case that if $\phi_* \lesssim 5.02$, which would have produced 60 efolds of inflation, the history will be allowable. The corresponding value of $f$, $(5.866\dots)$, is close to the numerical value $\approx 6.09$ that was obtained in \cite{Hertog:2023vot}.}
\label{naturalFIG}
\end{figure}

Here
\begin{equation}
	V = 1 + \cos(\phi/f) \,.
\end{equation}
The allowability-structure is similar to that of $\alpha$-attractors because we have a hilltop: there is a small region close to the end of inflation, $\phi_* \lesssim \chi \sim \pi f$, with an $\mathcal{O}(1)$ amount of efolds which are allowable, after which there is a non-allowable region, and then close to the hilltop near $\phi = 0$ trajectories become allowable once more. We find $f_{\textsf{KSW},60} = (5.866\dots)$ while \cite{Hertog:2023vot} found numerically $f_{\textsf{KSW},60} \approx 6.09$. We illustrate this in Figure \ref{naturalFIG}.

\section*{Acknowledgements}
We thank Victor Gorbenko, Thomas Hertog, Joel Karlsson and Juan Maldacena for inspiring conversations.

\appendix
\section{Slow roll corrections to the on-shell action} \label{OSAappendix}
The action is
\begin{align}
	S &= \int \di^4 x \sqrt{-g} \left( \frac{R}{2} - \frac{1}{2}\left( \partial \phi \right)^2 - V \right) + \text{(boundary)} \\
	&= 2\pi^2 \int \di t ~ 3a(1-\dot{a}^2) + a^3 \left( \frac{\dot{\phi}^2}{2} - V \right) \,.
\end{align}
Using the Friedmann equation \eqref{EOMfull1}, the on-shell action can be expressed as
\begin{equation} \label{Sbarintegral}
	\bar{S} = 4 \pi^2 \int \di t ~ a(3-a^2 V) \,.
\end{equation}
For de Sitter, $a(t) = (1/H_*) \cosh(H_* t)$, $V_* = 3 H_*^2$, and calculating the integral along the no-boundary contour from $t = i \pi / 2 H_*$ to $t = T$ on the real line where $a = b$, we find the classic result
\begin{equation}
	\frac{\bar{S}_\textsf{dS}}{4 \pi^2} = -\frac{3i}{V_*} - \frac{3}{V_*} \left[ (H_* b)^2 - 1 \right]^{3/2} \,.
\end{equation}
If as in \cite{Maldacena:2024uhs} we call $\alpha_r \equiv H_* b$ and expand at large $\alpha_r$:
\begin{equation}
	\frac{i\bar{S}_\textsf{dS}}{4 \pi^2} \times H_*^2 = 1 - i\left( \alpha_r^3 - \frac{3}{2} \alpha_r + \mathcal{O}(1/\alpha_r) \right) \quad \text{as } \alpha_r \to \infty \,.
\end{equation}
This agrees with Eq. (48) in \cite{Maldacena:2024uhs}. Now we would like to compute the leading slow roll correction to this. 

\subsection{Small field regime}
In the regime $\tau \ll 1/\sqrt{\varepsilon_*}$ we can use the analytic solution \eqref{EOMsol} of \cite{Maldacena:2024uhs}. We will fix the integration constants $c_{\phi,a}$ as \cite{Maldacena:2024uhs} does. In this way, with $N_e \equiv \tau - \log 2$,
\begin{equation}
	\varphi(\tau) \sim -N_e \,, \quad \gamma(\tau) \sim -\frac{N_e^2}{2} + \frac{N_e}{6} \quad \text{as } \tau \to \infty \,.
\end{equation}
We will calculate the integral in \eqref{Sbarintegral} from the south pole at $\tauSP = i \pi/2 - \varepsilon_* c_a$ to a point $\tau_\textsf{int}$ on the real line which satisfies $1 \ll \tau_\textsf{int} \ll 1/\sqrt{\varepsilon_*}$, keeping track of the leading corrections in $\varepsilon_*$ to the de Sitter answer. The point $\tauint$ must be chosen so that the same value $H_* b$ is attained at it as in the de Sitter solution. The relevant relation is
\begin{equation}
	H_* b = \alpha_r = \cosh \tauint \left( 1 + \varepsilon_* \gamma(\tauint) \right) \,.
\end{equation}
Inverting to first order in $\varepsilon_*$ we find\footnote{Notice the small imaginary contribution. This arises because we did not completely kill the imaginary part of $a(\tau)$ (or $\phi(\tau)$) on the real line with our choice of integration constants $c_{\phi,a}$ in \eqref{EOMsol}.}
\begin{equation}
	\tauint = \cosh^{-1} \alpha_r + \varepsilon_* \left[ \frac{1}{2} \log \alpha_r \left( \log \alpha_r - \frac{1}{3} \right) - \frac{3\log \alpha_r - 2}{4\alpha_r^2} - \frac{4i}{9 \alpha_r^3} + \mathcal{O}\left( \frac{\log \alpha_r}{\alpha_r^4} \right) \right] \quad \text{as } \alpha_r \to \infty \,.
\end{equation}
To first order in $\varepsilon_*$ we find\footnote{Because the integrand vanishes at the center of the ball, there is no contribution from the shift of the south pole at this order.}
\begin{align}
	\frac{\bar{S}_1}{4 \pi^2} = \frac{1}{H_*} \int_{\tauSP}^{\tauint} \di \tau \, a(3-a^2 V) &= \frac{\bar{S}_\textsf{dS}}{4 \pi^2} + \frac{3\varepsilon_*}{V_*} \int_{i \pi/2}^{\cosh^{-1}\alpha_r} \di \tau \left( 3 \gamma \cosh \tau - 9 \gamma \cosh^3 \tau - 6 \varphi \cosh^3 \tau \right) \\
	&+ \frac{3\varepsilon_*}{V_*} \left( \frac{1}{2} \log \alpha_r \left( \log \alpha_r - \frac{1}{3} \right) - \frac{3\log \alpha_r - 2}{4\alpha_r^2} - \frac{4i}{9 \alpha_r^3} + \cdots \right) 3\alpha_r(1-\alpha_r^2) \,.
\end{align}
The integral can be evaluated analytically; we do so and finally obtain
\begin{equation}
	\frac{i(\bar{S}_1 - \bar{S}_\textsf{dS})}{4 \pi^2} \times \frac{H_*^2}{\varepsilon_*} = i \alpha_r^3 \left( \log \alpha_r - \frac{1}{6} \right) + \frac{i \alpha_r}{4} \left( 6 \log \alpha_r - 11 \right) + i\pi + \left( \log 4 - \frac{7}{2} \right) + \mathcal{O}(1/\alpha_r) \,. \label{DeltaSsmalltau}
\end{equation}
This answer was obtained in \cite{Maldacena:2024uhs} using different coordinates.

\subsection{Large field regime}
A priori \eqref{DeltaSsmalltau} is valid in the regime $1 \ll \tauint \ll 1/\sqrt{\varepsilon_*}$, or $1 \ll \alpha_r \ll \exp \left( 1/\sqrt{\varepsilon_*} \right)$. Beyond this the approximation \eqref{EOMsol} we used breaks down as discussed in \S\ref{validitysec}. In \S\ref{largefieldsec} we discussed how the imaginary parts of the fields decay beyond this regime -- as long as inflation lasts -- and we can use this knowledge to estimate the contribution to the real part of $i \bar{S}$ coming from these late times. From \eqref{Sbarintegral} and \eqref{solutionX2} we have
\begin{align}
	\frac{1}{4\pi^2} \text{Im } \bar{S} \big|_{\textsf{late}} &= -\int_{\tint}^T \di t ~ 3 V(\phiRe) (\aRe)^2 \aIm + V'(\phiRe) (\aRe)^3 \phiIm + \text{(volume-suppressed)} \\
	&= -C \int_{\tint}^T \di t ~ \frac{V}{H} \left( 1 + \frac{1}{3} \left( \varepsilon_H + \eta_H \right) + \mathcal{O}(\varepsilon^2) \right) - \frac{V'}{\sqrt{2 \varepsilon_H}H} \left( 1 + \frac{2 \eta_H}{3} + + \mathcal{O}(\varepsilon^2) \right) \\
	&= -C \int_{\tint}^T \di t ~ H \, \mathcal{O}(\varepsilon^2) \,,
\end{align}
where we have used \eqref{EOMH2}. Notice the cancellation that has occurred at subleading order in slow roll. Without it, since $C = \mathcal{O}(\varepsilon_*/V_*)$, one might have expected a contribution $\mathcal{O}(\varepsilon_*/V_*)$ from this part of the evolution, since $\Delta \tau \sim 1/\varepsilon_*$ in large field models. This calculation shows the contribution is instead $\mathcal{O}(\varepsilon^2/V_*)$, where $``\varepsilon^2$" stands for second order in slow roll. So both the leading and first subleading contribution in slow roll to $\log |\Psi|^2$ originate from the approximately Euclidean (small field) part of the solution.

We have focussed here on potential subleading corrections to the real part of $i \bar{S}$ from the large field regime, which turn out to be absent. For a discussion of the leading behavior of the imaginary part of $i \bar{S}$ -- the phase of the wave function -- from the large field regime we refer the reader to \cite{Maldacena:2024uhs,Janssen:2020pii}.

\vfill
\bibliographystyle{klebphys2}
\bibliography{refs}
\end{document}